\documentstyle[12pt,epsfig,a4]{article}

\font\elevenrm=cmr10 scaled \magstep1

\def\Gcs{\hbox{GeV/$c^2$}}

\def\epem{\hbox{$\hbox{e}^+\hbox{e}^-$}}
\def\e{\hbox{e}}

\def\gaga{\hbox{$\gamma\gamma$}}
\def\nunu{\hbox{$\nu\bar\nu$}}
\def\lplm{\hbox{$\ell^+\ell^-$}}
\def\mumu{\hbox{$\mu^+\mu^-$}}
\def\toto{\hbox{$\tau^+\tau^-$}}
\def\ffb{\hbox{$\hbox{f}\,\overline{\hbox{f}}$}}
\def\ffbp{\hbox{$\hbox{f}\,\overline{\hbox{f}}\,'$}}
\def\qqb{\hbox{$\hbox{q}\overline{\hbox{q}}$}}

\def\g{\hbox{g}}

\def\d{\hbox{d}}

\def\cq{\hbox{c}}
\def\b{\hbox{b}}
\def\t{\hbox{t}}

\def\H{\hbox{H}}

\def\gas{\hbox{$\gamma^\ast$}}
\def\Z{\hbox{Z}}
\def\gZ{\hbox{$\gamma$/\hbox{Z}}}
\def\Zs{\hbox{$\hbox{Z}^\ast$}}
\def\W{\hbox{W}}

\def\Ws{\hbox{$\hbox{W}^\ast$}}

\def\WW{\hbox{$\hbox{W}^+\hbox{W}^-$}}

\def\mZ{\hbox{$m_{\hbox{\eightrm Z}}$}}

\def\tb{\hbox{$\tan\beta$}}

\def\inpb{\hbox{$\hbox{pb}^{-1}$}}

\def\st{\hbox{$\tilde{\hbox{t}}$}}

\def\slep{\hbox{$\tilde \ell$}}

\def\ser{\hbox{$\tilde{\hbox{e}}_R$}}

\def\smu{\hbox{$\tilde \mu$}}
\def\smr{\hbox{$\tilde \mu_R$}}

\def\stau{\hbox{$\tilde \tau$}}
\def\snu{\hbox{$\tilde\nu$}}
\def\grav{\hbox{$\tilde{\hbox{G}}$}}
\def\guno{\hbox{$\tilde{\hbox{g}}$}}

\def\chip{\hbox{$\chi^+$}}
\def\chim{\hbox{$\chi^-$}}

\def\mchi{\hbox{$m_\chi$}}
\def\mchip{\hbox{$m_{\chi'}$}}
\def\mchap{\hbox{$m_{\chi^+}$}}

\def\msmr{\hbox{$m_{\tilde\mu_R}$}}

\epsfverbosetrue
\def\mZ{\hbox{$m_{\rm Z}$}}

\title{SUPERSYMMETRIC PARTICLE SEARCHES AT LEP}
\author{J.-F. Grivaz\\
{\elevenrm Laboratoire de l'Acc\'el\'erateur Lin\'eaire,}\\ 
{\elevenrm IN2P3-CNRS et Universit\'e de Paris-Sud, F-91405 Orsay}}
\date{}%1 April 1996}

\begin{document}

\pagenumbering{arabic}
\pagestyle{plain}

\maketitle

\begin{picture}(160,1)
\put(130,85){\parbox[t]{45mm}{\bf LAL 97-61}}
\put(130,80){\parbox[t]{45mm}{August 1997}}
\end{picture}

\vfill

\begin{abstract}
Searches for supersymmetric particles performed at LEP~1 and LEP~2 are reviewed.
Using the MSSM with R-parity conservation as a reference model, the various 
analyses are briefly described, and the results are presented in terms of mass 
and coupling limits. Further implications of these results are discussed, 
including lower limits on the mass of a neutralino LSP, assuming the MSSM with 
GUT relations. Less conventional scenarii, among which those involving R-parity 
violation, are also investigated.
\end{abstract}
\vfill
\centerline{To appear in {\it Perspectives on Supersymmetry}}
\centerline{Gordon L. Kane, editor (World Scientific Publishing Company)}

\newpage 
 
\parskip = 3mm plus 1mm

\section{Introduction}

\subsection{The LEP history}

From 1989 to 1995, LEP, CERN's large \epem\ collider, operated at centre-of-mass
energies close to the Z mass. Each of the four experiments, ALEPH, DELPHI, L3
and OPAL, collected an integrated luminosity in excess of 150~\inpb,
corresponding to more than 4 million hadronic Z decays. Starting in the autumn 
of 1995, the beam energy was raised in steps with the adjunction of
superconducting RF cavities: the centre-of-mass energy reached 136 GeV at the
end of 1995, 161 GeV (just above the threshold for W pair production) in the
summer of 1996 and 172 GeV in the autumn of 1996. Integrated luminosities around
6, 11 and 11~\inpb\ were accumulated by each experiment at 130~--~136, 161 and
170~--~172~GeV, respectively. 
It is foreseen that the operation in 1997 will take place at an 
energy of 183~GeV, with further increases to 192~GeV in 1998 and possibly close 
to 200~GeV in 1999. In the following, the operation at and near the Z peak will
be referred to as LEP~1, the operation at 130~--~136~GeV as LEP~1.5, and the
operation at higher energies as LEP~2.

\subsection{General features of supersymmetric particle searches at \epem\ 
colliders}

Most of the searches for supersymmetric particles at \epem\ colliders are
inspired by the phenomenology of the Minimal Supersymmetric extension of the
Standard Model (MSSM), although their results often apply in a broader
framework. In particular, it is generally assumed that R-parity is conserved 
and that the Lightest Supersymmetric Particle (LSP) is the lightest 
neutralino~$\chi$. 
This leads to the celebrated signature of supersymmetry: missing energy. 

Since all charged supersymmetric particles are fairly democratically produced 
via $s$ channel \gZ\ exchange in \epem\ annihilation, the searches are 
naturally directed toward the Lightest Charged Supersymmetric Particle (LCSP), 
typically a scalar lepton or a chargino. This is in contrast to the situation 
at hadron colliders where only strongly interacting particles such as squarks 
or gluinos are abundantly produced. Moreover, in actual model calculations, it
usually turns out that cascades do not play an important role in the decay of 
the LCSP. The phenomenology is therefore rather straightforward, again in 
contrast to the situation at hadron colliders.
As an example, if the LCSP is the right smuon~\smr, 
the only production mechanism is $s$ channel \gZ\ exchange, 
with well defined couplings. The only decay mode is $\smr\to\mu\chi$ 
(at least if $\mchip>\msmr$), which leads to a final state consisting of 
a \mumu\ pair with missing energy. The absence of any signal in a given data 
sample is easily translated into a mass lower limit for the smuon.

This simple approach needs to be refined however, in particular when the 
production of neutral supersymmetric particles is taken into account. 
The lightest 
neutralinos or the sneutrinos can be pair produced via $s$ channel Z exchange, 
in which case they contribute to the invisible Z width, while the production 
of heavier neutralinos may lead to observable final states, possibly with an 
energy threshold lower than the one of LCSP pair production.
For instance, it is a rather common feature in supersymmetric models that
\hbox{$\mchi+\mchip<2\mchap$}, 
in which case $\epem\to\chi\chi'$ could be kinematically
allowed while $\epem\to\chip\chim$ is not. 
The value of the $\Z\chi\chi'$ coupling is however highly model dependent, and 
the search result usually cannot be translated into mass limits in a simple way.
This unpleasant feature is more than compensated by the fact that, at the 
expense of fairly simple and general hypotheses, the results from searches in 
different channels can be combined in a consistent way, sometimes leading to 
more powerful constraints than could have naively been expected.

\subsection{Specific features of the searches at LEP~1 and LEP~2}

With the steady increase of the LEP beam energy since 1995, it could be expected
that only the most recent results, namely those obtained at the highest 
energies, are relevant for supersymmetric particle searches. 
While this {\it a priori} belief is indeed found to be valid when comparing 
the results from LEP~2 and LEP~1.5, it does not fully hold when considering 
the LEP~1 results for two reasons. Firstly, the precision measurement of the 
Z boson properties, in particular of its total width, allow mass limits on
supersymmetric particles to be obtained independently of their decay patterns.
Secondly, the large statistics collected at LEP~1 allow Z decay branching 
ratios at the $10^{-6}$ level to be probed, which provides irreplaceable 
constraints in the neutralino sector.

As already mentioned, the main signature of supersymmetry is missing energy, at
least when R-parity is conserved. From this point of view, there are big
differences between LEP~1 and LEP~2. 
For processes such as chargino or slepton pair production, the kinematic limit 
of $\mZ/2$ had been reached at LEP~1 within months, with only a few thousand Z 
decays collected. The reason is that there is no irreducible background from
standard model processes: energy can be lost along the beam axis in \gaga\ 
interactions ($\epem\to\epem\ffb$) where the final state electrons escape
undetected in the beam pipe, or inside jets in the form of neutrinos from 
heavy flavour semileptonic decays in hadronic final states; both of these 
backgrounds can be eliminated by simple cuts on the direction and the isolation 
of the missing momentum. For processes such as Z decays into neutralinos, much
smaller signal to background ratios deserve attention so that rare effects have 
to be considered such as fake missing energy due to instrumental effects or 
wrong missing momentum direction due to a conspiration of missing energy
sources; this renders the analysis much more involved, but the techniques used 
are identical to those developed for the standard model Higgs boson 
search~\cite{PAJ} in the $\epem\to\H\nunu$ channel and they will not be detailed
here.

At LEP~2, on the contrary, new standard model processes take place which lead to
large missing energy in configurations much more difficult to disentangle from
those expected from signals of supersymmetry. Examples of such processes are W
pair production, with at least one $\W\to\ell\nu$ decay, ZZ or \Z\gas\ pair 
production, with $\Z\to\nunu$, or single W production in the process
$\epem\to\W\hbox{e}\nu$. It has nevertheless been possible to reduce these 
backgrounds to a level which remains negligible, at least with the modest 
integrated luminosities collected until now, without giving up too much signal 
efficiency, as will be explained in some detail further down. As the statistics 
accumulated increases, it may however become necessary to accept some level of
background; but the situation is better from this point of view than at LEP~1 in
the sense that this background is due to well calculable processes rather than 
to less controllable instrumental effects.

\subsection{Synopsis and warnings}

The rest of this chapter is organized as follows. The conventional scenario, 
with R-parity conservation and with a neutralino LSP, will be discussed first, 
starting with the basic facts, namely the mass and coupling limits obtained at 
LEP~1 and LEP~2, and then proceeding toward the interpretation in the MSSM, 
with in particular the derivation of mass limits for the LSP. Less conventional 
scenarii will be considered next, involving for instance a light gravitino or 
\hbox{R-parity} violation. Finally, a brief outlook toward the future will be 
given. 

Unless otherwise stated, all limits are given at 95\% confidence level (CL).
The results quoted are, as much as possible, extracted from published papers. 
However, when no publication was available on a given topic at the time of 
writing, preliminary results submitted to the Summer '97 conferences, and to 
which the author had access, have been used. 
There has been no attempt toward a complete list of references. Only those 
from which the quoted results have been extracted are explicitly given.
Moreover, it has not been judged useful to give theoretical references since 
the necessary theoretical background can be found in this book. 
Finally, although a major prediction of supersymmetry is the existence of a 
light Higgs boson, which could well lie within the reach of LEP~2, no discussion
of this issue is presented here since the searches for supersymmetric Higgs 
bosons at LEP are described in detail in Ref.~[1].

\section{The conventional scenario: basic facts}

In the conventional scenario, R-parity is conserved and the lightest
supersymmetric particle is colourless and electrically neutral. This leaves the
gravitino, a sneutrino or the lightest neutralino as possible candidates, which 
implies in turn that the LSP is weakly interacting with ordinary matter, hence 
the missing energy signature of supersymmetry. The conventional choice for the
LSP is the lightest neutralino $\chi$.

Although most of the searches for supersymmetric particles at LEP have been 
conducted using the MSSM as a reference model, the results obtained are often 
fairly general. It is the purpose of this section to review these basic facts.

\subsection{Constraints from the Z width}

One of the principal goals of the LEP~1 run, and in particular of the scans
performed in the vicinity of the Z peak, was the precise determination of the 
parameters of the Z resonance: mass, total and partial decay widths, cross
section at the peak. In the standard model, these parameters can be accurately 
computed as a function of the mass of the top quark and, to a lesser extent, 
of the Higgs boson, these heavy particles contributing via virtual effects,
in particular in the Z boson propagator. This allowed the top quark mass to be 
predicted where it was finally measured, and now gives indications in
favour of a light Higgs boson, a feature in agreement with the expectation from
supersymmetry. Contributions from heavy supersymmetric particles to the Z
parameters are however too small to allow interesting constraints to be obtained
in a similar way.

On the other hand, if supersymmetric particles (or any kind of new particles) 
are light enough to be produced in Z decays, they will increase, often in a 
significant fashion, the total Z width with respect to the standard model
expectation. Therefore, the agreement between the predicted and
measured Z widths allows constraints on supersymmetric particles kinematically
accessible in Z decays to be inferred. The Z width is now measured to
be~\cite{EW} $2494.7 \pm 2.6$~MeV. The prediction is 2502.0~MeV, for a top mass
of 175~\Gcs, a Higgs mass of 150~\Gcs, and a value of 0.118 for the strong 
coupling constant $\alpha_S$. This prediction is decreased to 2498.4~MeV,
allowing for values as low as 0.115 for $\alpha_S$ and 169~\Gcs\ for the top
mass, corresponding to a one standard deviation change for each~\cite{Warsaw}.
The value of 150~\Gcs\ for the Higgs mass is an upper limit in supersymmetry, 
so no further decrease of the predicted Z width is possible from this origin. 
This leaves a maximum of 3.4~MeV for any contribution to the Z width due to 
supersymmetric particle production (at 95\% CL, using Bayesian statistics). 
This corresponds to 2\% of the partial width $\Gamma_\nu$ of the Z into a single
flavour of \nunu\ pair.

Very light charginos would contribute $4.5 \Gamma_\nu$ if pure gauginos, or 
$\sim 0.5 \Gamma_\nu$ if pure higgsinos. This is large enough to exclude
charginos practically up to \mZ/2, irrespective of their field content.
This limit is much lower than the typical ones achieved nowadays at LEP~2, but 
it applies independently of the chargino decay pattern and remains the only
one to be valid in some extreme configurations, as will be discussed further
down.

Light sneutrinos would contribute $0.5 \Gamma_\nu$ for each flavour. Because the
phase space factor is less favourable than for charginos, this results into a
limit of only 43~\Gcs\ for a single flavour. For three mass degenerate flavours,
the limit is again half the Z mass. Constraints on an invisible sneutrino,
applicable if it is the LSP or even if the decay mode $\snu\to\nu\chi$ is
dominant, can be inferred in a similar way from the measurement of the Z partial
width into invisible final states (or equivalently from the effective number of
neutrinos). In contrast to the situation of a few years ago, they turn out to 
be hardly stronger than those inferred from the total width measurement.

Somewhat weaker limits can also be derived from the Z width for sleptons or 
squarks. In the case of neutralinos, the couplings to the Z are highly model 
dependent (and even parameter dependent within a given model). The coupling is
largest for higgsino-like neutralinos, and vanishes for pure gauginos. The best 
that can be done is therefore to set coupling (or branching ratio) upper limits 
as a function of the involved neutralino masses. However, these are superseded 
by those obtained from direct searches, except in the $\Z\to\chi\chi$ case
where the final state is invisible.

\subsection{Constraints from direct searches at LEP~1}

As discussed just above, it is only in the case of neutralinos that direct 
searches at LEP~1 play a major role. 
The most relevant channels leading to visible final states are 
$\Z\to\chi\chi'$, kinematically the most favourable, and to a lesser extent 
$\Z\to\chi'\chi'$. 
The main $\chi'$ decay mode is $\chi'\to\chi\ffb$ which is accessed 
through virtual Z or sfermion exchange ($\chi'\to\chi\Zs$; 
$\chi'\to\hbox{$\tilde{\hbox{f}}^\ast\,\overline{\hbox{f}}$}$). 
In some particular 
instances the $\chi'\to\chi\gamma$ mode, which proceeds via loops, can become 
significant or even dominant. (This happens for small $\chi'$--$\chi$ mass
differences or if one of the neutralinos is almost purely higgsino and the 
other one almost purely photino.) The possible final states resulting from 
$\chi\chi'$ production are therefore: purely invisible ($\chi\chi\nunu$), 
lepton pairs ($\chi\chi\lplm$), jets ($\chi\chi\qqb$) or single photons 
($\chi\chi\gamma$) with missing energy.

At LEP~1, the only significant standard sources of lepton or jet pairs are Z 
decays and \gaga\ interactions. In the first case there is no missing energy, 
ignoring for the moment neutrinos involved in $\tau$ or heavy flavoured hadron 
decays. In \gaga\ interactions, on the contrary, the spectator electrons 
({\it i.e.} the electrons which radiated the photons participating in 
the collision) tend to remain undetected in the beam pipe, giving rise to a 
large amount of missing energy. The direction of the missing momentum is 
however close to the beam axis and there is only little missing transverse 
momentum $p_T$. In both Z decays and \gaga\ interactions, the leptons or the 
jets therefore appear back-to-back in the plane transverse to the beam axis. 
Even for $\tau$ pairs, this (almost) coplanar topology is preserved in the 
visible decay products, and the same holds in the case of semileptonic decays 
of heavy flavoured hadrons. This feature is the basis of all searches for 
``acoplanar'' leptons or jets. Of course, the detectors should be as hermetic 
as possible to ensure that no additional particles, for instance photons 
radiated at large angle, escape detection, thus inducing fake missing $p_T$.
In the case of single photons, radiative Bhabha events ($\epem\to\epem\gamma$)
with both electrons remaining undetected in the beam pipe can be eliminated by a
cut on the transverse momentum of the photon corresponding to the maximum $p_T$
that these two electrons can carry without entering the detector acceptance. The
same cut also rejects events from $\epem\to\gaga\gamma$, with two photons close 
to the beam axis. 
The actual analyses performed by the LEP experiments follow those general
principles, but they ended up being appreciably more involved both because the
detectors are not ideal and because standard physics is more complicated. 
(For instance, while a two-jet event with missing energy in one of the jets 
remains coplanar, this is no longer the case for a three-jet event.)

\begin{figure}[p]
\epsfig{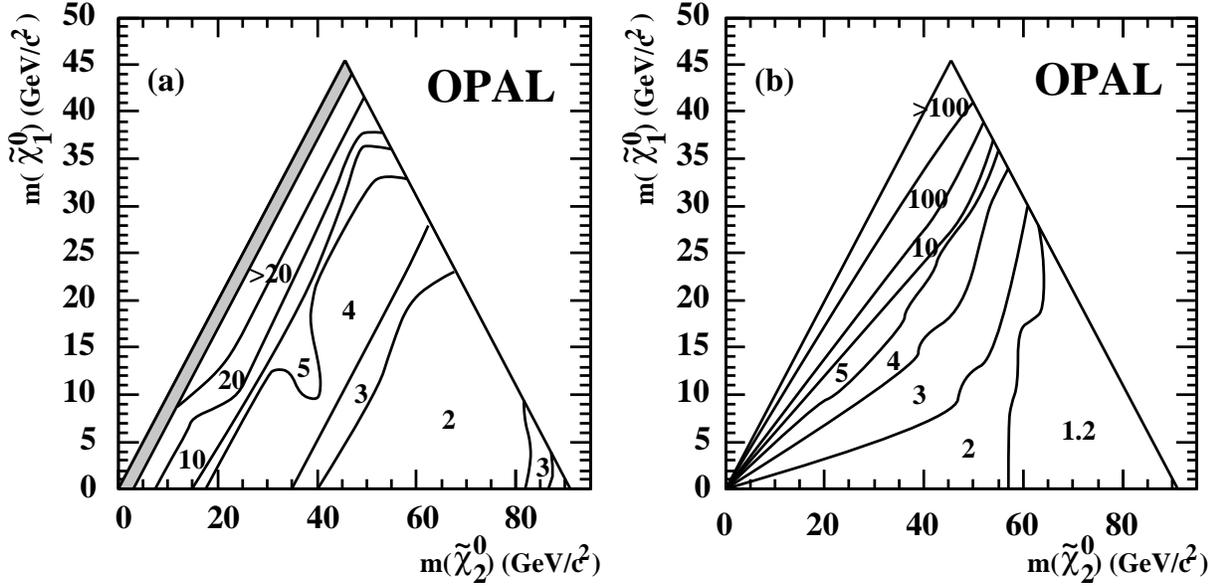}
\caption{Upper limits on product branching ratios times $10^6$ for neutralino
production:\break
(left) $\hbox{BR}(\Z\to\chi'\chi)\hbox{BR}(\chi'\to\chi\Zs)$;
(right) $\hbox{BR}(\Z\to\chi'\chi)\hbox{BR}(\chi'\to\chi\gamma)$. 
\label{fig:ochi}}
\end{figure}

Events from irreducible standard model backgrounds are expected to be selected, 
but at a very low level. Indeed, a few spectacular ``monojet'' events were 
found~\cite{monoj},
but they can be explained with not unreasonably small probabilities as
originating from four fermion final states such as \lplm\nunu\ or \qqb\nunu, 
reached through the process $\epem\to\Zs\gas$, with $\Zs\to\nunu$ and
$\gas\to\lplm\hbox{ or }\qqb$. Such an event is shown in 
Fig.~\ref{fig:monojtoto}(top).
Similarly, no single photon signal was observed 
beyond the background expected from $\epem\to\Zs\gamma$. 
With the full LEP~1 statistic, upper limits at the level of a few $10^{-6}$ 
have thus been set~\cite{ochi} for the product branching 
ratios $\hbox{BR}(\Z\to\chi'\chi)\hbox{BR}(\chi'\to\chi\Zs)$ and
$\hbox{BR}(\Z\to\chi'\chi)\hbox{BR}(\chi'\to\chi\gamma)$, as shown in 
Fig.~\ref{fig:ochi}.

\subsection{Results from the searches at LEP~2}

The searches for supersymmetric particles performed at LEP~2 address sleptons, 
stops, charginos and neutralinos. Sneutrinos are expected to decay invisibly 
($\snu\to\nu\chi$) and cannot be searched for efficiently. 
Gluinos are not produced directly in \epem\ collisions and have 
therefore not been considered. In contrast to the case of generic squarks,
for which the limits obtained at the Tevatron cannot be rivaled at LEP, large 
mixing can be expected in the stop sector, as will be discussed further down. 
This may lead to a top squark significantly lighter than all other squarks, 
hence the relevance of stop searches at LEP.

\subsubsection{Sleptons and stops}

Sleptons are pair produced and have been searched in the decay mode
$\slep\to\ell\chi$, {\it i.e.} in final states consisting of acoplanar lepton
pairs of the same flavour. Similarly, pair produced stops have been searched in
the acoplanar jet topology expected to arise from the $\st\to\cq\chi$ decay
mode. (The normal decay mode, $\st\to\t\chi$, is kinematically forbidden.) 
This effectively flavour changing neutral current process occurs at the
one loop level, and the corresponding decay width is small enough for the top
squark to hadronize into a stop hadron before decaying, a feature which has
been implemented by the LEP collaborations in their Monte Carlo generators. 
In the particular case where the sneutrino is lighter than the stop, the
$\st\to\b\ell\snu$ decay mode is expected to become dominant. This rather
peculiar mass hierarchy has also been addressed at LEP, but it will not be 
considered further here.

A number of backgrounds to these acoplanar lepton and acoplanar jet searches 
have already been discussed in the context of neutralino searches at LEP~1. 
Furthermore, abundant fermion pair production occurs through radiative return to
the Z ($\epem\to\gamma\Z\to\gamma\ffb$), leading mostly to final states with 
large missing energy along the beam axis. 
All these essentially coplanar backgrounds are reduced 
at LEP~2 in a way similar to that at LEP~1. In addition, new backgrounds arise 
from processes such as $\epem\to\WW$,~$\W\e\nu$,~$\Z\e\e$ or~$\Z\Zs$ which lead
to four-fermion final states with missing energy originating from leptonic 
decays such as $\W\to\ell\nu$ or $\Z\to\nunu$ or from electrons escaping
undetected in the beam pipe in the case of $\W\e\nu$ or $\Z\e\e$. Although these
backgrounds are irreducible at some point, their kinematic properties are
ne\-vertheless sufficiently different from those expected from slepton or stop 
pair production to allow almost background free samples to be selected with the 
presently accumulated integrated luminosities. In the case of smuon searches for
instance, requiring two leptons identified as muons selects only 1\% of the W
pairs. Moreover, the typical momentum of muons from smuon decay is smaller 
than that of muons from W decay, a feature which becomes more and more
pronounced as the neutralino mass increases. In the case of acoplanar jets, the
background from \WW\ with $\W\to\ell\nu$ is reduced by a veto against energetic 
leptons, and even against isolated particles to cope with $\W\to\tau\nu$ decays.
An additional requirement that the mass of the visible system should not exceed 
some value slightly lower than the W mass also helps in reducing the $\W\e\nu$
background. In the end, a few events are selected, compatible with the 
background from standard model processes, typically \WW\ with $\W\to\tau\nu$, 
or \Z\Zs\ with $\Z\to\nunu$. An example of an event probably due to this last 
process, with $\Zs\to\toto$, is shown in Fig.~\ref{fig:monojtoto}(bottom).

Almost model independent mass limits can be derived for smuons and staus which 
are produced only via $s$ channel \gZ\ exchange, with well defined couplings, 
under the assumption that $\smu\to\mu\chi$ and $\stau\to\tau\chi$ are the only 
decay modes. (In the MSSM, this assumption is usually valid for right sleptons 
within the mass range of interest here, given the other constraints existing 
in the chargino/neutralino sector.) At the time of writing, a \smr\ mass limit 
of 59~\Gcs\ is obtained~\cite{aslep} for \smr--$\chi$ mass differences
exceeding 10~\Gcs. In the case of selectrons, $t$ channel neutralino exchange
also contributes, usually increasing the production cross section. But this
contribution is model dependent and no fully general result for selectrons can 
be extracted. In Fig.~\ref{fig:seleslep}(left), 
an example of \ser\ mass limit is
presented~\cite{aslep}, valid in the MSSM for the set of parameters indicated.
(Here, account has been taken of decay modes other than $\ser\to\e\chi$.)

It is worth mentioning at this point that results from lower energy machines in
this sector can still be relevant. The single photon final state has been 
studied at PEP, PETRA and TRISTAN. In the standard model, it originates through
initial state radiation from the reaction $\epem\to\gamma\nunu$. 
A supersymmetric contribution could arise in a similar way from the process 
$\epem\to\gamma\chi\chi$, mediated by $t$ channel selectron exchange. The
absence of any excess above the standard model expectation allows selectrons
with masses up to 79~\Gcs\ to be excluded (90\% CL) in the case of an almost 
massless and photino-like LSP~\cite{AMY}.

\begin{figure}[ht]
\epsfig{file=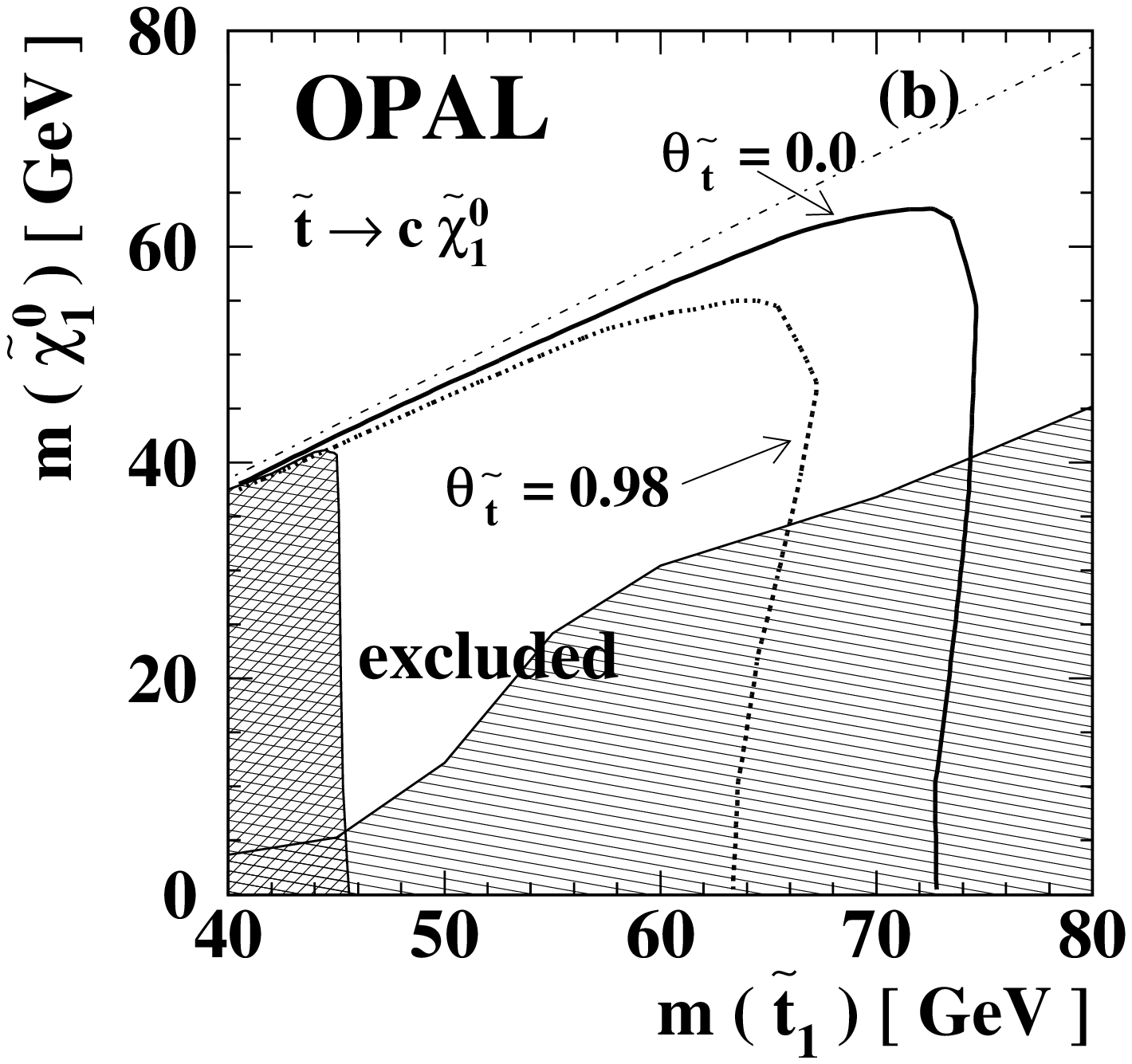,width=8.6cm,
bbllx=0pt,bblly=0pt,bburx=450pt,bbury=450pt}\hfill
\epsfig{file=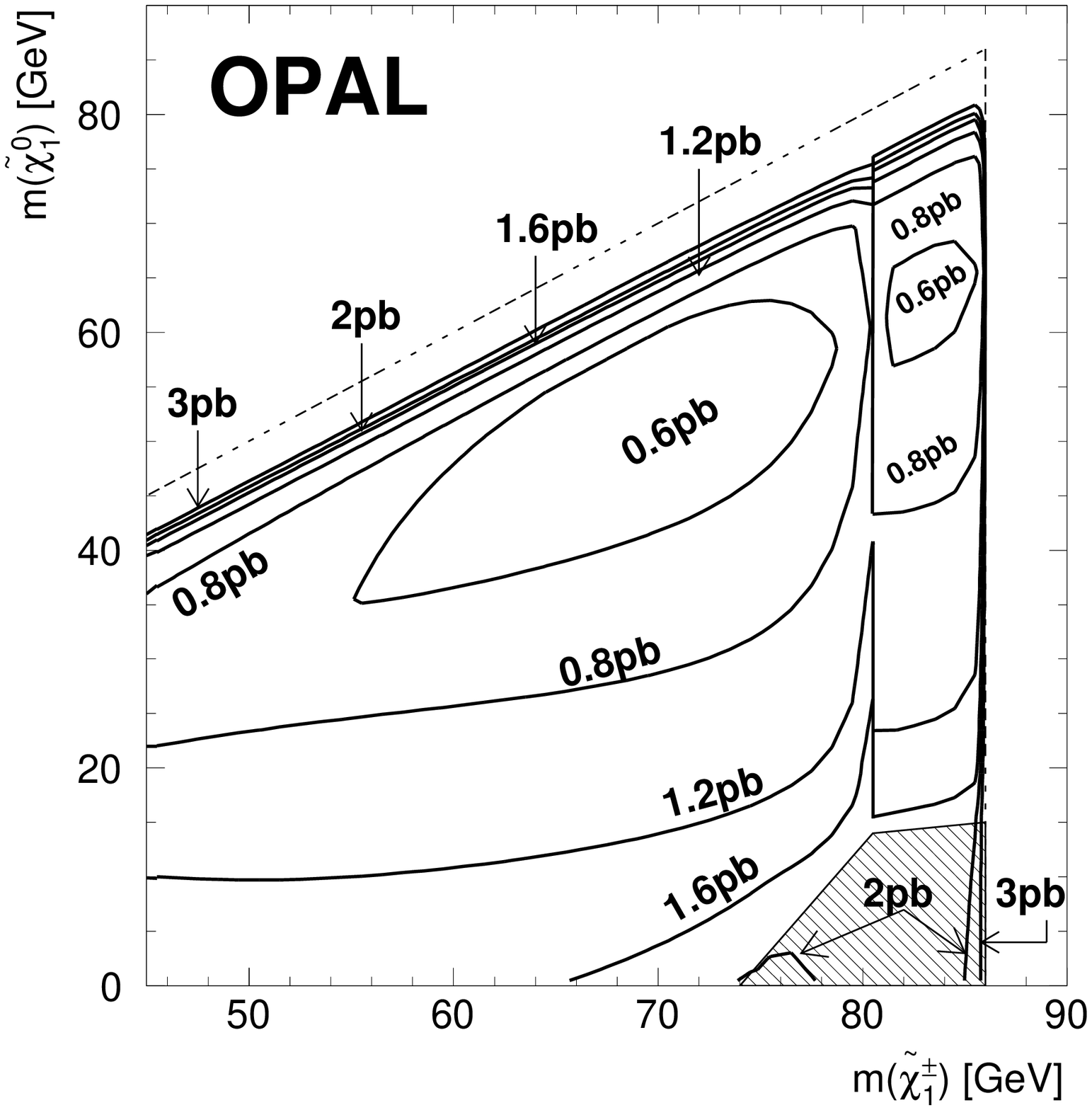,width=7.1cm,
bbllx=0pt,bblly=175pt,bburx=550pt,bbury=705pt}
\caption{(left): Excluded regions in the plane of the \st\ and $\chi$ masses, 
for two values of the mixing angle in the stop sector, 0.0 and 0.98 rad
corresponding to the most and least favourable cases, respectively. The cross
hatched region has been excluded at LEP~1. The singly hatched region has been
excluded by D0 at the Tevatron.
(right): Cross section upper limit for chargino pair production at 172~GeV in 
the plane of the \chip\ and $\chi$ masses.
\label{fig:stopchar}}
\end{figure}

In the case of stops, the mass limit depends on the value of the mixing angle 
in the stop sector which controls the coupling of the stop to the Z. This
coupling is largest for a pure left stop (null mixing angle) and vanishes for a
value of the mixing angle close to 0.98 radians. The results~\cite{ostop} for
these two extreme cases are shown in Fig.~\ref{fig:stopchar}(left). 
The exclusion domain 
obtained at the Tevatron is extended toward smaller \st--$\chi$ mass 
differences.

\subsubsection{Charginos and neutralinos}

Pair production of charginos is mediated by $s$ channel \gZ\ exchange, and by
$t$ channel sneutrino exchange, with destructive interference between the two
processes. The influence of sneutrino exchange is largest for light sneutrinos
and for gaugino-like charginos. 
The main chargino decay mode is $\chip\to\chi\ffbp$ which is accessed 
through virtual W or sfermion exchange ($\chip\to\chi\Ws$; 
$\chip\to\hbox{$\tilde{\hbox{f}}^\ast\,\overline{\hbox{f}}\,'$}$). For
gaugino-like charginos and sleptons significantly lighter than squarks, the
latter contribution enhances leptonic decays. If sneutrinos are sufficiently
light, two-body decays open up ($\chip\to\ell^+\snu$). 
The possible final states resulting from chargino pair production are
therefore: acoplanar leptons ($\chi\nu\ell^+\chi\bar\nu\ell^-$ 
or $\snu\ell^+\tilde{\bar\nu}\ell^-$), purely hadronic with missing energy 
($\chi\hbox{q}\,\overline{\hbox{q}}\,'\chi\hbox{q}\,'\overline{\hbox{q}}\,$)
or mixed ($\chi\nu\ell\chi\hbox{q}\,\overline{\hbox{q}}\,'$).

The searches for acoplanar leptons from charginos are similar to those designed 
for slepton pairs, with minor differences: the two leptons need not be of the
same flavour, but their momenta are softer than in the case of sleptons.
With four primary quarks, the topology of the purely hadronic final states is
not identical to the acoplanar jet topology encountered in the case of stop pair
production, which again induces modifications to the analyses: the visible
mass cut is somewhat relaxed, but the events are required to exhibit a more 
spherical pattern. Undoubtedly, the gold plated topology is the mixed one: with
an isolated lepton and missing energy in a hadronic environment, the backgrounds
from fermion pair production and from \gaga\ interactions are easy to reduce to
a negligible level; the {\it a priori} harmful background from
$\epem\to\WW\to\ell\nu\hbox{q}\,\overline{\hbox{q}}\,'$ 
can be distinguished from
the signal by the mass of the hadronic system (close to the W mass for the \WW\ 
background, smaller for a signal in the mass reach of LEP~2) and by the missing
mass (close to zero for the \WW\ background, significantly larger for the
signal, a feature which is enhanced as the $\chi$ mass increases). 
In practice, the signal topology and the background conditions are also 
affected by the \chip--$\chi$ mass difference which controls the amount of
visible energy: the \gaga\ background and the trigger conditions are of greater
concern for small mass differences, while the \WW\ background is most dangerous
for large mass differences. Therefore, the LEP collaborations were led to 
design matrices of analyses according to the type of final state and to the 
mass difference. 

Small numbers of events were selected by the various experiments, but at a level
compatible with the expectation from standard model processes. In the absence 
of signal, results on chargino production have been expressed in an almost 
model independent way in terms of cross section upper limit as a function of the
\chip\ and $\chi$ masses, assuming dominance of the decay through virtual W
exchange ({\it i.e.} heavy sleptons and squarks). (The ``almost'' restriction
comes from the fact that, for a given \chip\ mass, the decay kinematics depend
not only on the \chip--$\chi$ mass difference but also, however to a much 
lesser extent, on the detailed \chip\ and $\chi$ field content.) 
%At the time of writing, only preliminary results were available from the 1996 
%run at 172~GeV, an example~\cite{achar} of which is given 
An example of such limits~\cite{ochar} is given in 
Fig.~\ref{fig:stopchar}(right). 
Further specification of the model is needed to translate these results into
chargino mass limits. The statement can however be made in a fairly general way
that the kinematic limit of 86~\Gcs\ is practically reached for gaugino-like 
charginos when all sfermion masses are very large.

Neutralino production, $\epem\to\chi_i\chi_j$ where $\chi_i$ and $\chi_j$ stand
for any of the neutralinos $\chi$, $\chi'$, $\chi''$,\dots proceeds via $s$
channel Z exchange and via $t$ channel selectron exchange. In contrast to
chargino production, the interference is normally constructive. The process with
the lowest threshold leading to visible final states is $\epem\to\chi\chi'$.
As already discussed in the context of LEP~1, acoplanar lepton pairs and
acoplanar jets are the relevant topologies. The analyses are thus similar
to those set up to search for sleptons or stops. Further selections were
developed for the other processes, involving multileptons or isolated photons. 
The range of application of these selections depends strongly on the field 
contents of the various neutralinos involved, as they control both the 
production cross sections and the decay branching ratios, including cascade 
decays. It is therefore difficult to express the results in a way at the same 
time general and meaningful. In the absence of any signal up to a centre of 
mass energy of 172~GeV, the discussion of the implications of the searches for 
neutralinos is deferred to the next section.

\section{The conventional scenario: interpretation}

As indicated a number of times in the previous section, the results obtained by
the various LEP collaborations usually cannot be translated directly into 
general supersymmetric particle mass limits; further specification of the model 
is needed for that purpose. After specifying the theoretical framework commonly
referred to under the logo MSSM by the LEP collaborations, the basic facts 
presented in the previous section will be turned into constraints on the 
parameters of this model, and it will be shown how the combination of various 
inputs allows mass lower limits on the lightest neutralino to be inferred.  

\subsection{Theoretical framework}

In the Minimal Supersymmetric Extension of the Standard Model (MSSM), the gauge
group of the minimal standard model is assumed, and only the minimal field 
content is introduced. In particular, there are just two Higgs doublets, and 
hence four neutralinos. The parameters needed to specify the model are a set of 
soft supersymmetry breaking masses and trilinear couplings, $m_i$ and $A_i$, 
for all  scalar doublets and singlets, and of soft supersymmetry breaking
masses, $M_i$, for the three gauginos. In the Higgs sector, a supersymmetric 
Higgs mass term, $\mu$, is introduced, and the ratio $v_2/v_1$ of the vacuum 
expectation values developed by the Higgs fields coupling to the up-type and 
down-type quarks is denoted \tb.

Additional simplifying assumptions are commonly made, inspired by Supergravity
(SUGRA) models: for all matter sfermions, a universal scalar mass $m_0$ and a 
universal trilinear coupling $A$ at the Grand Unification (GUT) scale; 
and for the three gauginos, a universal mass $m_{1/2}$ at the GUT scale. 
The low energy parameters are then determined using the renormalization 
group equations. The gaugino masses evolve in the same way as the gauge 
coupling constants. As a result, the gluino is expected to be much heavier, 
typically 3.5 times, than the lighter chargino in large regions of the
parameter space. 
It is also expected that sleptons are lighter than squarks, and right sleptons 
or squarks lighter than their left counterparts; the sneutrino mass should be 
similar to the left slepton mass, unless \tb\ is large in which case 
sneutrinos become lighter. Mixing among the left and right scalars is expected 
to be small, except in the stop sector because of the large mass of the top 
quark, and possibly for staus if \tb\ is large.

In ``Minimal SUGRA'', the hypothesis of universal scalar masses is extended 
to the Higgs sector, and the spontaneous breaking of the electroweak
symmetry is triggered at low energy by radiative corrections to the Higgs
masses. Imposing the proper mass for the Z boson reduces the set of parameters
further: the $\mu$ parameter is determined from the others, up to its sign.

\subsection{Experimental results}

\begin{figure}[ht]
\epsfig{file=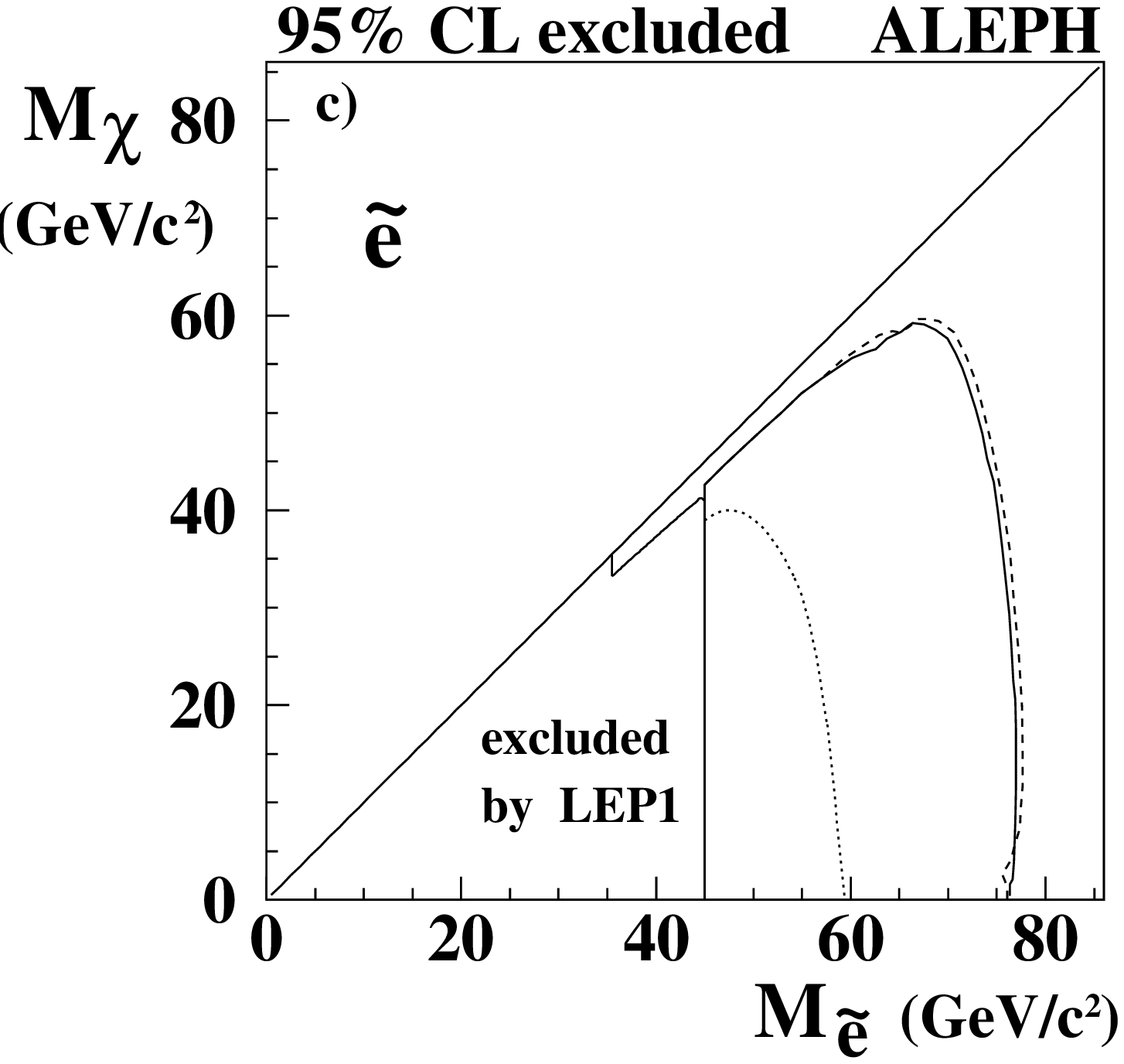,width=8cm,
bbllx=0pt,bblly=0pt,bburx=470pt,bbury=440pt}\hfill
\epsfig{file=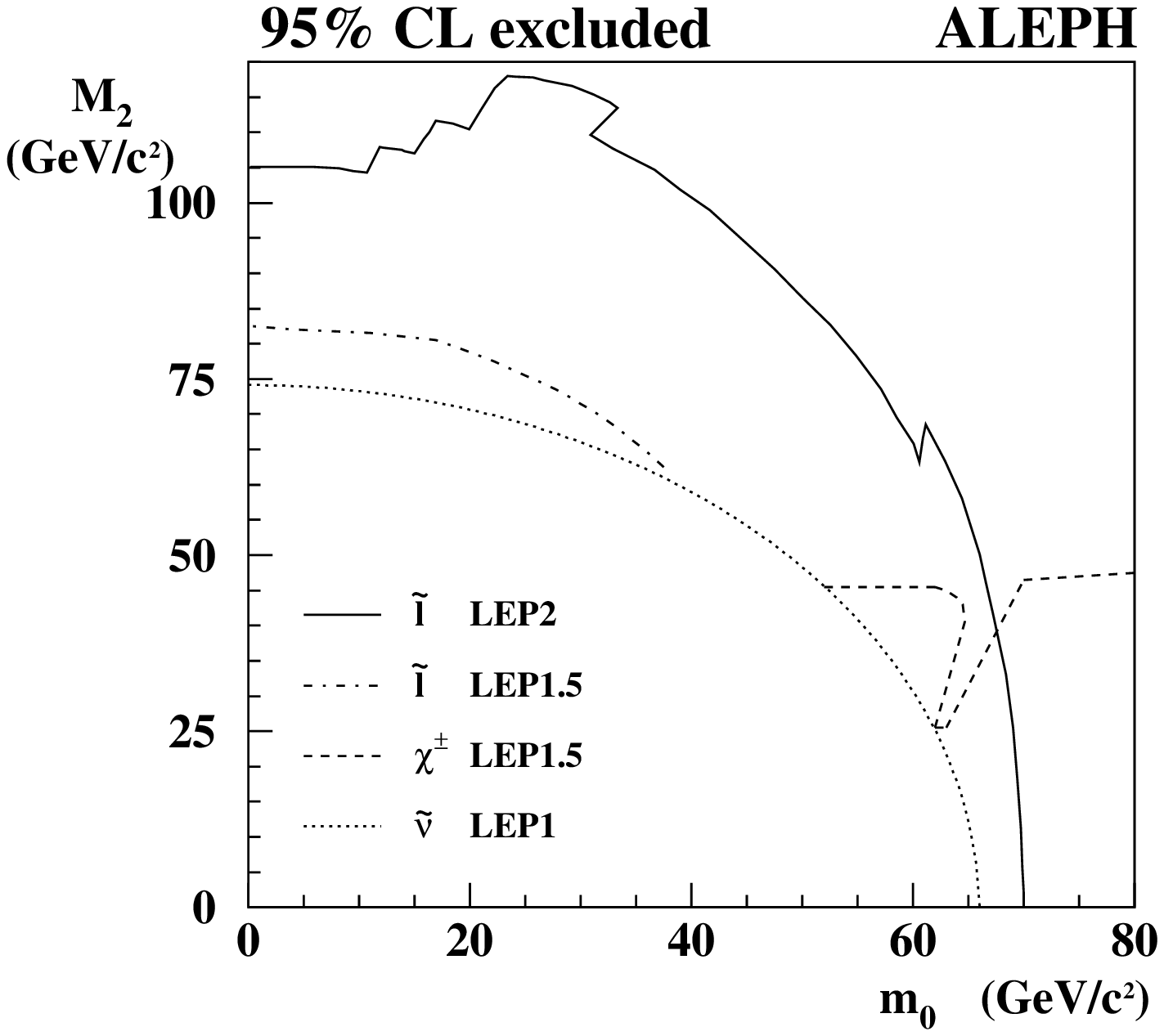,width=8cm,
bbllx=0pt,bblly=0pt,bburx=460pt,bbury=425pt}
%\vspace{0.4cm}
\caption{(left): Excluded regions in the plane of the \ser\ and $\chi$ masses, 
in the MSSM with $\tb=2$ and for $\mu=-200$~\Gcs\ (solid) and $\mu=1000$~\Gcs\
(dashed). The dotted curve is the LEP1.5 limit.
(right): Excluded regions in the ($m_0$,$M_2$) plane, for $\tb=2$ and
$\mu=-200$~\Gcs\ (solid curve). 
\label{fig:seleslep}}
\end{figure}

An example of results obtained within the framework of the MSSM has already been
given for right selectrons in Fig.~\ref{fig:seleslep}(left). 
Here, The GUT relation among
gaugino masses has been assumed to calculate the contributions of the various 
neutralino exchanges in the production process and the branching ratio for the
decay $\ser\to\e\chi$. The results are presented for a specific value of \tb, 
and for two representative choices of $\mu$. The searches for all slepton 
flavours can be further combined assuming scalar mass universality, and turned 
into an exclusion region~\cite{aslep} in the ($m_0$, $M_2$) plane, as shown in 
Fig.~\ref{fig:seleslep}(right). 
($M_2$ and $m_{1/2}$ are related by $M_2 = 0.82 m_{1/2}$.)
Again, this exclusion is valid for the specified \tb\ and $\mu$ values.

\begin{figure}[ht]
\epsfig{file=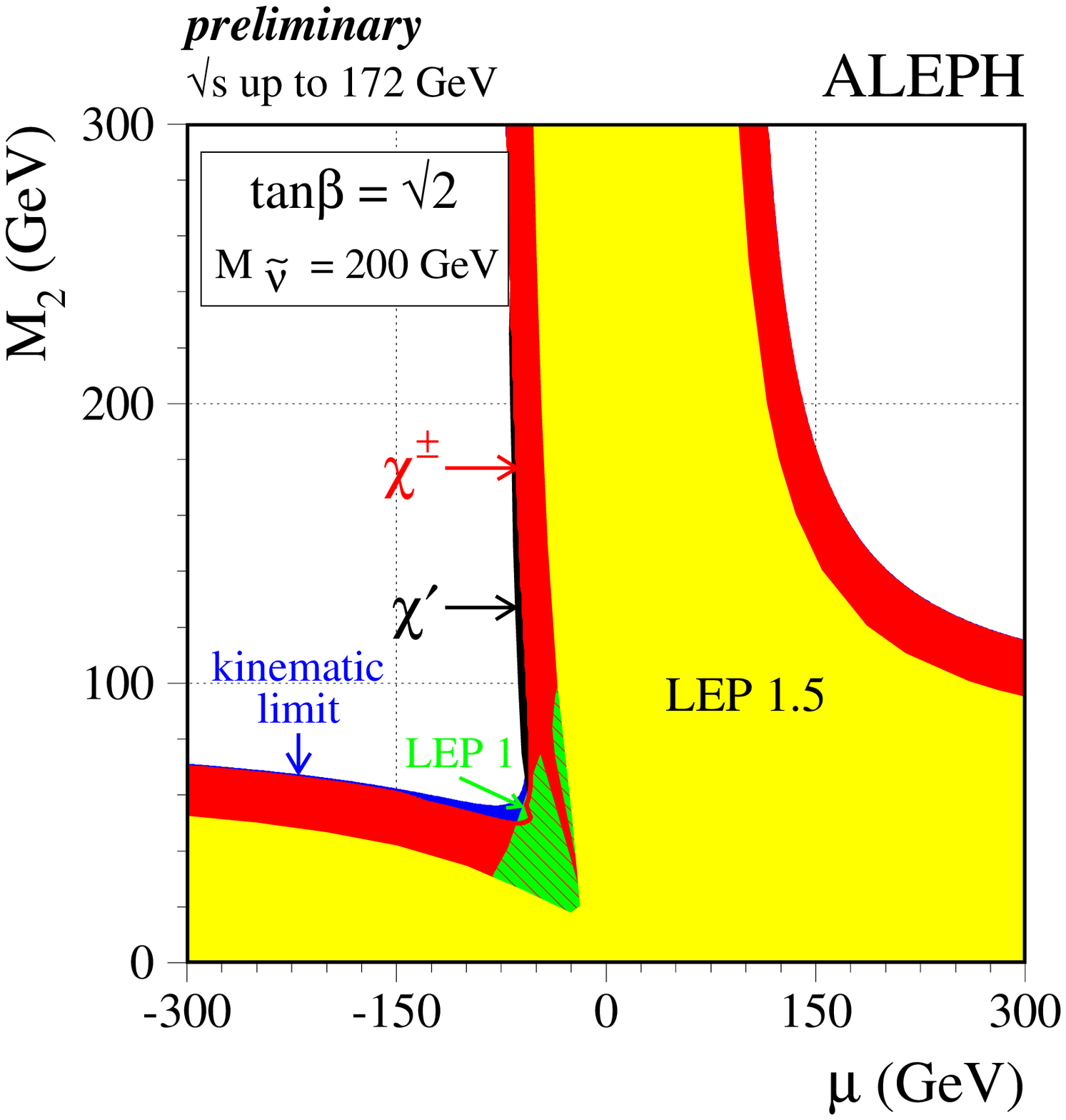,width=7.9cm,
bbllx=0pt,bblly=145pt,bburx=495pt,bbury=675pt}\hfill
\epsfig{file=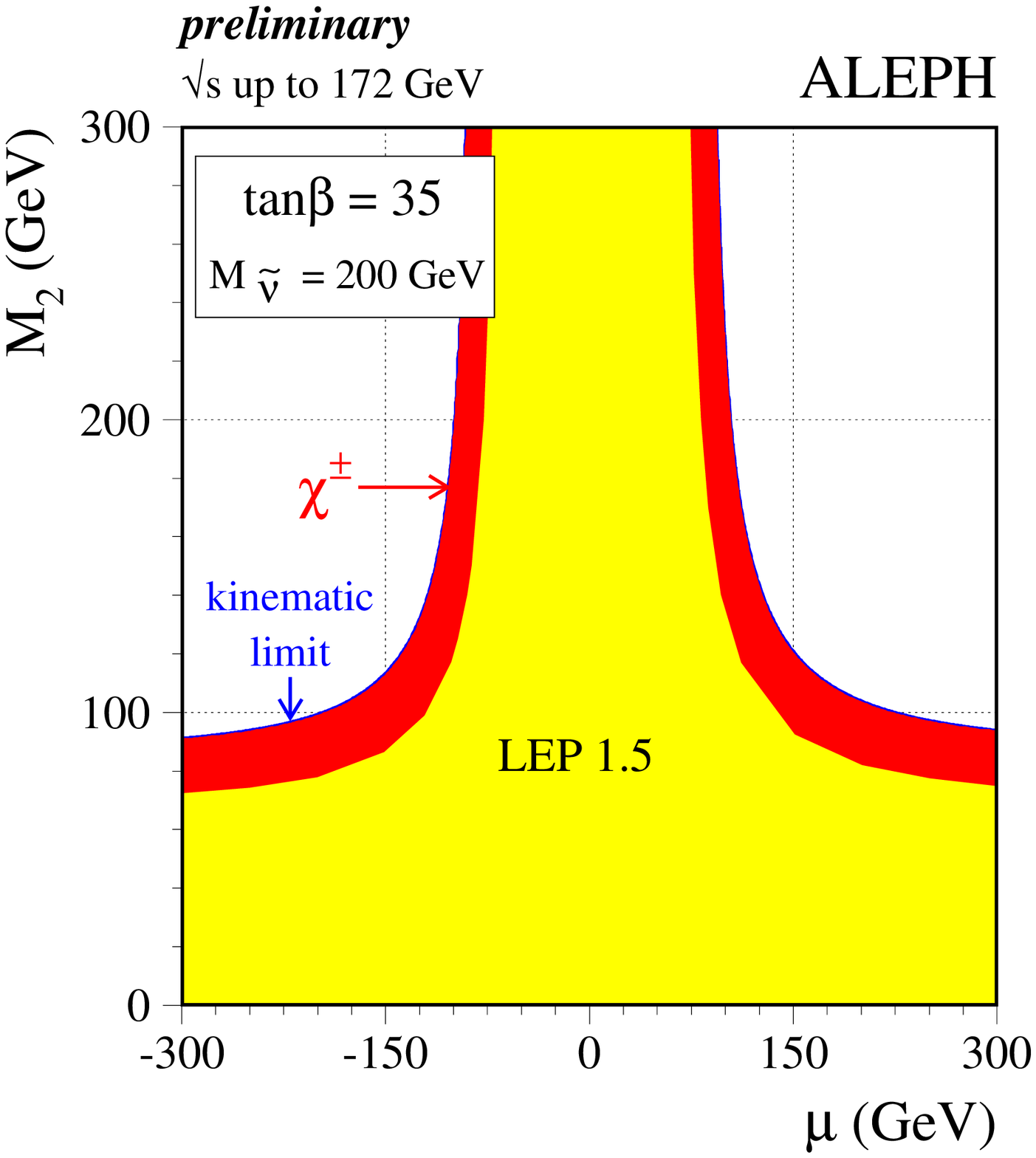,width=7.1cm,
bbllx=0pt,bblly=120pt,bburx=495pt,bbury=650pt}
\caption{Excluded regions in the ($M_2$,$\mu$) plane for $\tb=1.41$ (left) 
and $\tb=35$ (right), assuming heavy sfermions ($m_0=200$~\Gcs).
\label{fig:mmu}}
\end{figure}

Results on chargino production such as those shown in 
Fig.~\ref{fig:stopchar}(right) 
are
commonly translated into exclusion domains in the ($M_2$, $\mu$) plane, for
chosen values of \tb. Since the cross section limits depend not only on the 
chargino mass but also on the mass of the lightest neutralino, the GUT relation 
between $M_1$ and $M_2$ has to be assumed. This allows in turn account to be
taken of possible reductions in the detection efficiency due to cascade decays 
such as $\chip\to\chi'\Zs$. The analysis is simplest with the assumption that 
all sfermions are sufficiently heavy not to affect the production cross section 
nor the decay branching ratios. 
Exclusion domains~\cite{achar} determined in this way 
%in case from cross section limits such as those shown in 
%Fig.~\ref{fig:stopchar}(right) 
are presented in Fig.~\ref{fig:mmu} for two values of \tb, 1.41 and 35. 
(Such values are typical of the low and large \tb\ solutions in the so-called 
infra-red quasi fixed point scenario.) With the same hypotheses, all
neutralino production cross sections and decay branching ratios can be
calculated from the same set of parameters ($M_2$, $\mu$ and \tb, still with the
assumption of heavy sfermions). The additional exclusion domains resulting from 
the neutralino searches are also displayed in Fig.~\ref{fig:mmu}.

\begin{figure}[ht]
\epsfig{file=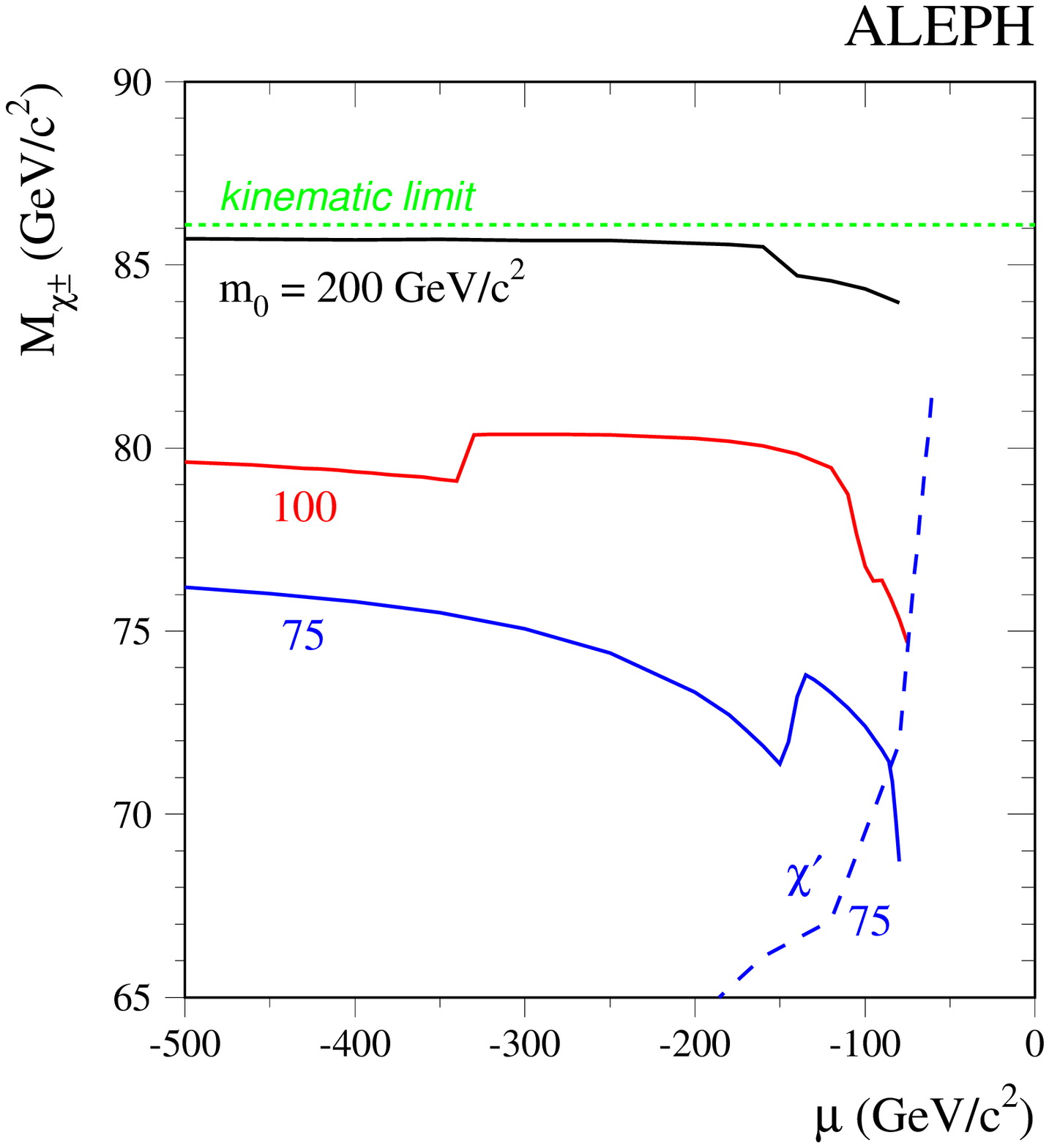,width=7.1cm,
bbllx=0pt,bblly=110pt,bburx=500pt,bbury=650pt}\hfill
\epsfig{file=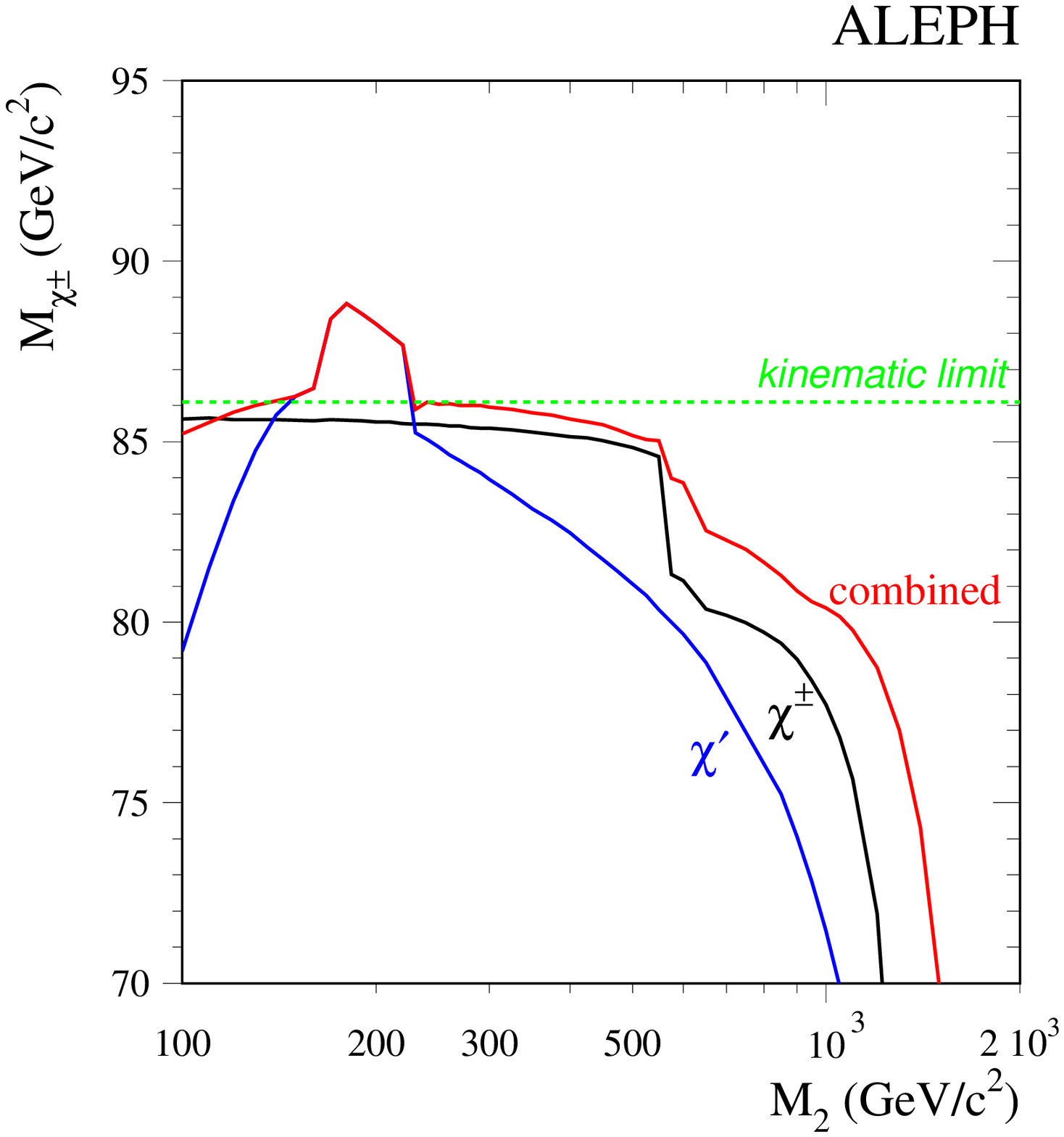,width=7.1cm,
bbllx=0pt,bblly=110pt,bburx=500pt,bbury=650pt}
\caption{Chargino mass limit as a function of $\mu$ (left) and of $M_2$ (right),
for $\tb=1.41$. On the left, the limits from the chargino searches are given for
various values of $m_0$, and the dashed curve is the limit from the neutralino
searches for $m_0 = 75$~\Gcs. On the right, the limits from the chargino,
neutralino and combined searches are given for $m_0 = 200$~\Gcs.
\label{fig:cham}}
\end{figure}

These exclusion contours can in turn be translated into chargino mass limits, as
shown in Fig.~\ref{fig:cham} where the limit is displayed as a function of $\mu$
for gaugino-like charginos and as a function of $M_2$ for higgsino-like 
charginos. The decrease in the limit for very large values of $M_2$ is due to
the fact that the \chip--$\chi$ mass difference, and hence the selection
efficiency, diminishes as $M_2$ increases. The impact of the neutralino
searches is clearly visible for moderate values of $M_2$, allowing chargino
masses to be indirectly excluded beyond the kinematic limit.

These results are modified for lower sfermion masses. Light sneutrinos reduce
the production cross section for gaugino-like charginos, while light selectrons
increase the production cross section for gaugino-like neutralinos. Since also
the decay branching ratios are affected by the sfermion masses, the assumption
of a universal scalar mass $m_0$ is made so that the various production and 
decay processes are correlated. The influence of light sfermions is shown in 
Fig.~\ref{fig:cham}(left). 
For $m_0 = 75$~\Gcs, for instance, it can be seen that 
the chargino mass limit is reduced by $\sim 10$~\Gcs, and it would degrade even 
further for $\mu>-100$~\Gcs\ if the constraints from the neutralino searches 
were not taken into account.

Altogether, even if no hard number can be given as a lower mass limit for
charginos, masses smaller than 75 to 85~\Gcs\ are excluded over most of the
parameter space. There remains however a loophole when the sneutrino mass is
very close to but smaller than the chargino mass. In such a case, the two-body 
decay $\chip\to\snu\ell^+$ dominates, but the final state lepton has too little 
energy and the detection efficiency vanishes. There is nothing to be recovered
from the neutralino searches because the dominant decay mode, $\chi'\to\snu\nu$,
is invisible, and the only limit remaining is the one deduced from the Z width
measurement which is insensitive to the chargino decay pattern.

\subsection{Neutralino LSP mass limits}

Direct searches for the lightest neutralino $\chi$ cannot be performed
efficiently at LEP~2. The relevant process is $\epem\to\chi\chi$, leading to an
invisible final state. At energies well below the Z peak, initial state
radiation tagging ($\epem\to\gamma\chi\chi$) had been used, leading to
correlated constraints on the neutralino and selectron masses, as already
mentioned in the context of selectron mass limits. This technique is however
useless at LEP~2 energies because of the overwhelming background from
$\epem\to\gamma\nunu$.

\begin{figure}[ht]
\epsfig{file=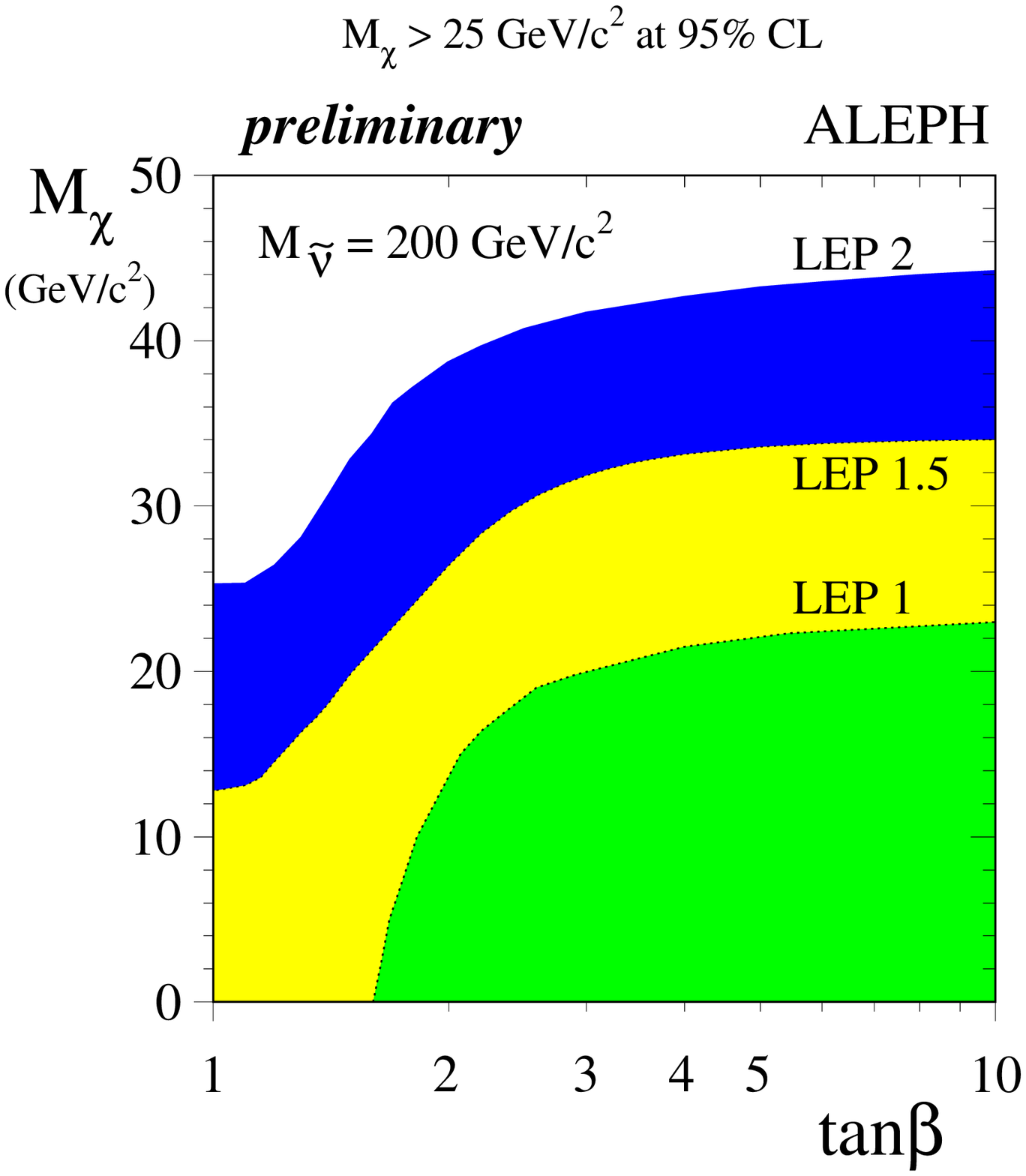,width=7.1cm,
bbllx=0pt,bblly=25pt,bburx=500pt,bbury=615pt}\hfill
\epsfig{file=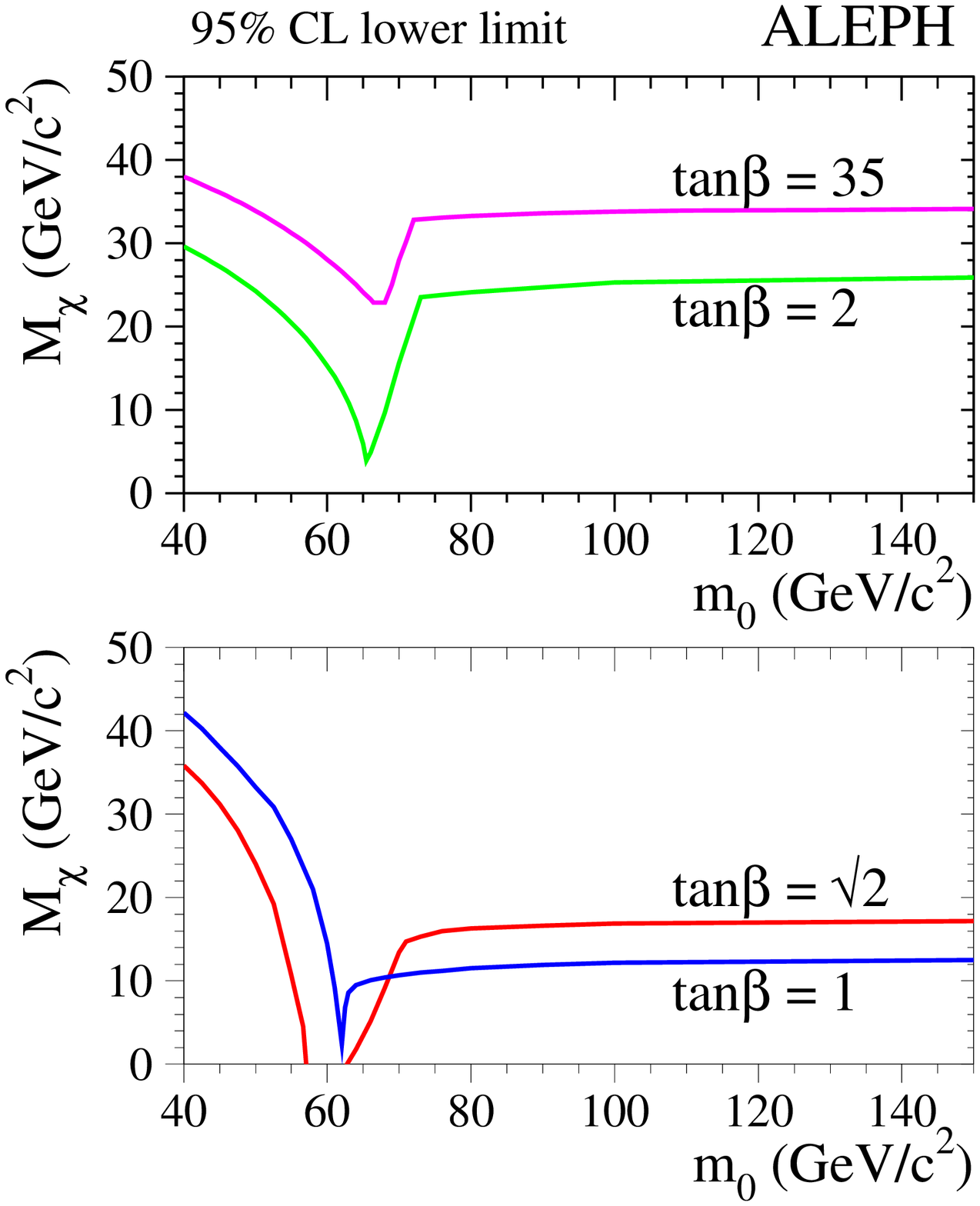,width=7.1cm,
bbllx=0pt,bblly=35pt,bburx=500pt,bbury=650pt}
\caption{(left) Mass limit for the lightest neutralino as a function of \tb\ 
and for heavy sfermions ($m_0 = 200$~\Gcs).
(right): Mass limit obtained at LEP~1.5 for the lightest neutralino as a 
function of $m_0$ and for various values of \tb\ ($\mu<0$). 
\label{fig:mchichim}}
\end{figure}

However, once the model is sufficiently specified, indirect limits can be
obtained from the constraints on charginos and on the more massive neutralinos.
For instance, by scanning the domain remaining unexcluded in Fig.~\ref{fig:mmu},
no set of ($M_2$, $\mu$) value can be found for which the mass \mchi\ of the 
lightest neutralino is smaller than 31~\Gcs, for $\tb=1.41$ and assuming 
heavy sfermions ($m_0$ = 200~\Gcs). It is worth noticing that, in this 
particular example, the LEP~1 exclusion still plays some role, due to the fact
that the neutralino search is limited, in the region near the point where the 
$\chi$ mass limit is set, by the value of the couplings to the Z rather than by
kinematics. (In a general fashion, the LEP~1 constraints remain useful for low
values of \tb.) The lower limit obtained~\cite{arrgh} for \mchi\ 
as a function of \tb\ and for heavy sfermions ($m_0 = 200$~\Gcs)
is displayed in Fig.~\ref{fig:mchichim}(left).

For lower sfermion masses, as already discussed, the assumption of a universal 
scalar mass $m_0$ is needed to render the analysis manageable. Since, as has 
been seen above, the constraints in the chargino/neutralino sector are weaker 
in that case, it is not surprising that the limits on \mchi\ also degrade. For
$m_0 = 75$~\Gcs\ and $\tb = 1.41$, the $\chi$ mass lower limit is 21~\Gcs\
(instead of 31~\Gcs\ for $m_0 = 200$~\Gcs). At the time of writing, a 
complete analysis, letting both $m_0$ and \tb\ vary freely, is not available
at LEP~2. The results obtained~\cite{avant} at LEP~1.5 will therefore be used 
in the following for the purpose of illustration of the methods. 

In Fig.~\ref{fig:mchichim}(right), 
the limit obtained for \mchi\ is displayed, for various 
values of \tb, as a function of $m_0$. For large values of $m_0$, they reflect
the equivalent of Fig.~\ref{fig:mchichim}(left) 
at LEP~1.5. As $m_0$ decreases, the limit
degrades slowly, due to the destructive interference in the chargino production
cross section for light sneutrinos. As $m_0$ decreases further, the sneutrino
mass becomes smaller than the chargino mass and the detection efficiency
vanishes abruptly, as mentioned previously. However, for very low values of
$m_0$, the results from slepton searches, such as the LEP~1.5 counterpart of 
Fig.~\ref{fig:seleslep}(right), 
can be used (including the sneutrino mass limit from LEP~1 
which is most relevant for large values of \tb). In general, the exclusions 
from sleptons and from charginos overlap, and a massless neutralino is excluded 
irrespective of $m_0$. This is however not the case for $\tb = 1.41$ and 
$m_0 \sim 60$~\Gcs, as can be seen in Fig.~\ref{fig:mchichim}(right). 
Preliminary results~\cite{arrgh}
from LEP~2 indicate that this small region is now excluded, with a neutralino
LSP mass limit of 14~\Gcs, independent of $m_0$ and of \tb.

\begin{figure}[ht]
\epsfig{file=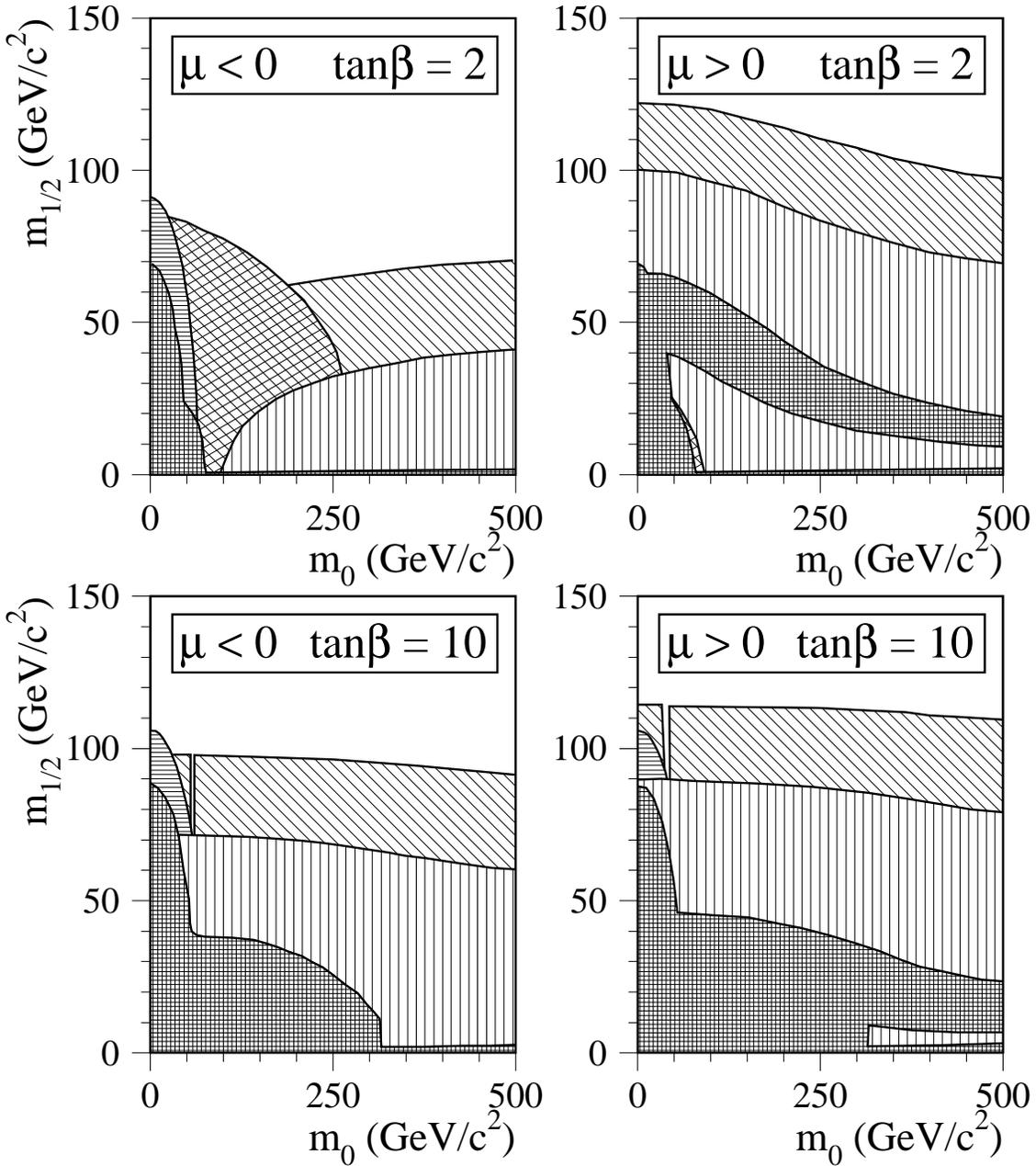,width=15.5cm,
bbllx=10pt,bblly=65pt,bburx=510pt,bbury=660pt}
\caption{Excluded domains in the ($m_0$,$m_{1/2}$) plane for $\tb=2$ (top) and
for $\tb=10$ (bottom), and for
$\mu<0$ (left) and $\mu>0$ (right). The dark shaded regions are theoretically
excluded. The vertical,horizontal,
crossed and slanted hatched regions are excluded by
charginos at LEP~1, by sneutrinos at LEP~1, by Higgs bosons at LEP~1 and by
charginos at LEP~1.5, respectively.
\label{fig:sugra}}
\end{figure}

More constraining results can be derived by imposing further 
constraints on the model. 
If minimal SUGRA is assumed, the $\mu$ value, up to now a free parameter,
can be calculated up to its sign from $m_0$, $m_{1/2}$ and \tb. Moreover, the 
results from the searches for Higgs bosons can also be used, particularly 
relevant for low values of \tb\ and for negative $\mu$. 
Examples of exclusion domains~\cite{avant} 
in the ($m_0$,$m_{1/2}$) plane are shown in Fig.~\ref{fig:sugra}. 
A $\chi$ mass lower limit of
22~\Gcs\ can be inferred this way from the LEP~1.5 data, independent of $m_0$, 
of \tb\ and of the sign of $\mu$. (Here too, this rather low value for the limit
on \mchi\ is entirely due to the loophole discussed earlier.)

\section{Unconventional scenarii}

\subsection{Constraints from LEP~1 on light gluinos}

In most of the analyses presented up to now, it has been assumed that gluinos
are too heavy to affect the decay pattern of the supersymmetric particles
considered. Indeed, stringent mass limits for gluinos have been obtained at
hadron colliders, as reported elsewhere in this book. Although some simplifying
assumptions had to be made when deriving these limits, for instance regarding 
the details of cascade decays, it is hard to see how relaxing these
assumptions could invalidate the exclusion of gluinos in the mass range of
interest at LEP, at least within the MSSM with GUT relations. 

There remains however a small mass window for very light gluinos (approximately 
from 2.5 to 4~\Gcs, for squarks in the few hundred \Gcs\ range) not officially 
excluded by any of the searches at hadron colliders or elsewhere ({\it e.g.} in 
beam dump experiments or in upsilon decays). Such light gluinos could invalidate
some of the searches performed at LEP~2. For instance, the dominant chargino 
decay mode could well be $\chip\to\qqb\guno$, with subsequent hadronization of
the gluino into a long-lived R-hadron. In that case, the final state from 
chargino pair production would not exhibit the characteristic signature of 
missing energy.

Such light gluinos would however modify the usual phenomenology of  QCD. 
For instance, they would affect the topology of four-jet events, 
via $\g\to\guno\guno$ splitting. More importantly, they would contribute 
to the running of $\alpha_S$ as three additional flavours, in leading order, 
up to mass effects. With the large sample of hadronic events collected at
LEP~1, such detailed studies have been performed~\cite{colfac}, excluding 
gluinos with masses smaller than 6.3~\Gcs.

\subsection{Stable charged particles}

In the conventional scenario, the LSP is assumed to be neutral and colourless,
based on cosmological arguments. Even in that case, it could nevertheless occur
that the LCSP is hardly heavier than the LSP, which could make it long lived 
without conflicting with cosmology. In the MSSM, this may occur for instance 
if the GUT relation between $M_1$ and $M_2$ is not satisfied. In scenarii with 
a light gravitino LSP, of which some implications are discussed further down, a 
slepton could be the next to lightest supersymmetric particle; for large enough 
values of the supersymmetry breaking scale, this slepton would decay with a 
lifetime long enough to appear as stable in a LEP detector.

Searches for long-lived weakly interacting heavy charged particles have been 
performed up to 172~GeV centre-of-mass energy~\cite{stable}. Pair production of 
such particles would resemble muon pair production, with a back to back topology
but with smaller particle velocities and hence larger specific ionization loss
($\d E/\d x$). The absence of any signal excludes charginos almost up to the 
kinematic limit of 86~\Gcs, for heavy sneutrinos, and right smuons or staus up 
to 67~\Gcs. (The selectron production cross section is affected by the details 
of the neutralino sector.)

\subsection{The light gravitino LSP scenario}

A class of supersymmetric models in which the gravitino is the lightest 
supersymmetric particle has recently received renewed attention, as discussed
elsewhere in this book. For LEP, the main feature of such models is that the 
lightest neutralino is expected to decay into a photon and a gravitino, 
$\chi\to\gamma\grav$. For practical purposes, the neutralino lifetime is 
negligibly small as soon as the gravitino mass is smaller than a few~eV/$c^2$. 

Pair production of the lightest neutralinos, $\epem\to\chi\chi$, then leads to 
a final state consisting of two acoplanar photons with missing energy. The 
background from the process $\epem\to\nunu\gaga$ is reduced by the requirement 
that the mass recoiling to the two photons should not be close to the Z mass; 
in addition, a minimum energy for both photons is required, the value of which 
depends on the $\chi$ mass considered. In the absence of signal, upper
limits~\cite{lphot} are set on the production cross section of $\chi$ pairs at 
the level of 0.25--0.35~pb. To turn these constraints into a $\chi$ mass limit, 
further specification of the model is required (right and left selectron 
masses, $\chi$ field content). Typically, values around 70~\Gcs\ are obtained.

\subsection{R-parity violation}

R-parity conservation was introduced to forbid lepton number or baryon number 
violating terms in the superpotential, otherwise allowed by supersymmetry and 
by gauge invariance. These terms are of the form 
$\lambda_{ijk}L_iL_j\overline{E}_k$, $\lambda'_{ijk}L_iQ_j\overline{D}_k$ or
$\lambda''_{ijk}\overline{U}_i\overline{D}_j\overline{D}_k$, where $i,j,k$ are
generation indices, $L$ and $Q$ are doublet superfields, and $U$, $D$ and $E$ 
are singlet superfields. The simultaneous presence of the last two types of term
would induce unacceptably fast proton decay. There is however no compelling 
theoretical justification for this prescription, and there exist alternatives, 
in which only some subsets of these terms are present, and which are 
equally viable. In particular, it is sufficient to assume that only one (or more
generally only one type) of the R-parity violating terms is present to obtain 
an acceptable phenomenology. 

The main consequence of R-parity violation is that the LSP is no longer stable.
For instance, a $\lambda_{ijk}L_iL_j\overline{E}_k$ term may induce the 
decay of a neutralino LSP to final states such as $\bar\nu_i\ell^+_j\ell^-_k$
(via virtual slepton or sneutrino exchange) or of a sneutrino $\tilde\nu_i$ to 
a lepton pair $\ell^-_j\ell^+_k$. The characteristic signature of supersymmetry 
therefore no longer consists in missing energy, but rather in multileptonic 
final states. Searches for supersymmetric particles have been performed at 
LEP~1~\cite{marie} and LEP~2~\cite{rviol} under the assumption that R-parity is 
violated by such a $\lambda_{ijk}L_iL_j\overline{E}_k$ term. A large variety of 
final state topologies arise, depending on the type of particles produced
in the \epem\ annihilation, and on whether they decay directly via an R-parity
violating interaction or first toward the LSP via a gauge interaction (with the 
LSP subsequently decaying as indicated above in the case of a neutralino LSP). 
The generation indices in the $\lambda_{ijk}$ Yukawa coupling control the lepton
flavours appearing in the final state and therefore affect the selection 
efficiencies; the most favourable case is obtained for $\{ijk\}=\{122\}$, with 
only electrons and muons as final state leptons, and the worst case for 
$\{ijk\}=\{133\}$, with taus also appearing in the final state. 

\begin{figure}[ht]
\epsfig{file=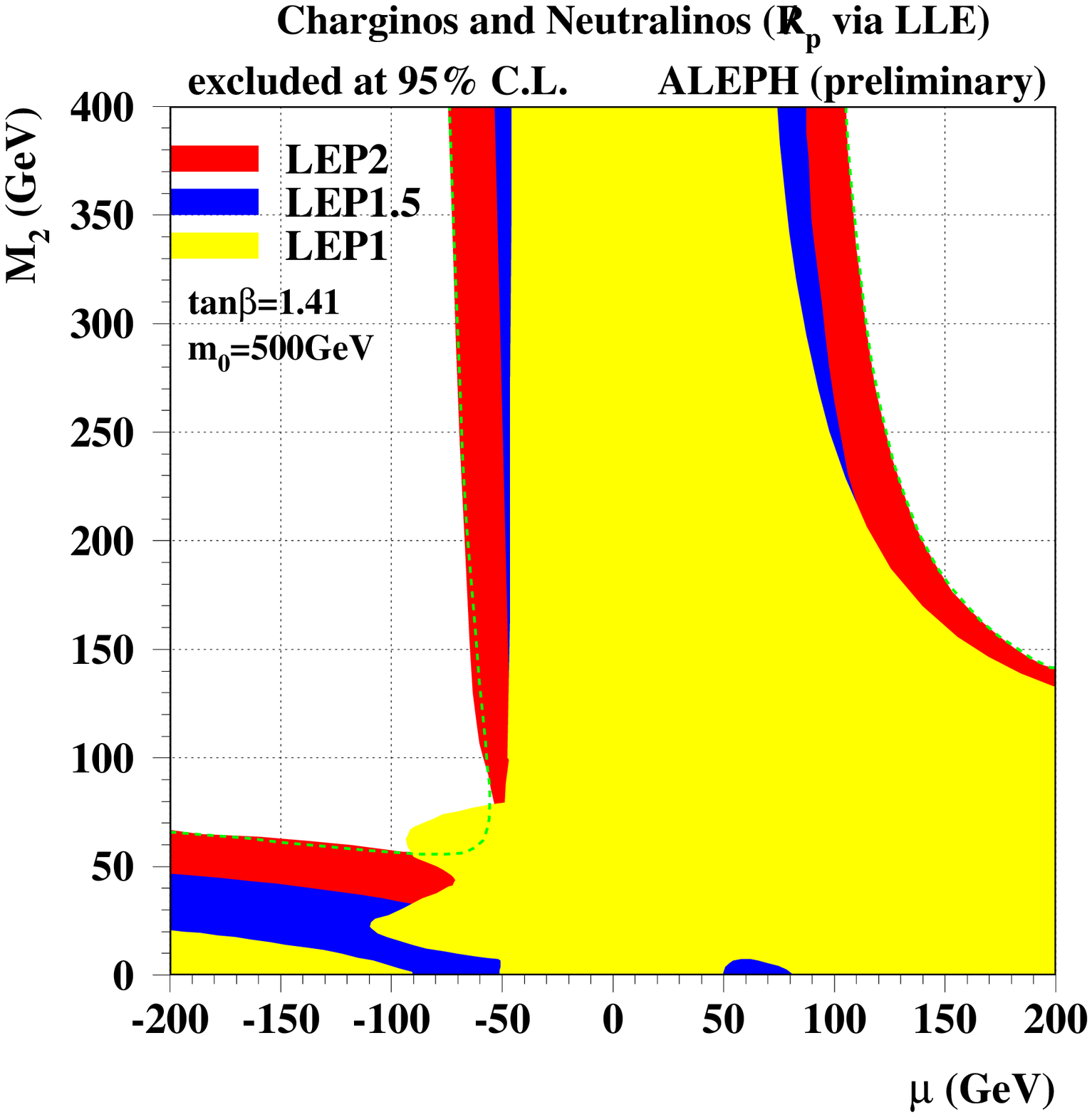,width=12.1cm,
bbllx=-115pt,bblly=140pt,bburx=470pt,bbury=710pt}
\caption{Excluded regions in the ($M_2$,$\mu$) plane for $\tb=1.41$ and for 
heavy sfermions, 
assuming that R-parity is violated by a $\lambda_{ijk}L_iL_j\overline{E}_k$
term. 
The dashed line is the kinematic limit for $\mchap=86$~\Gcs.
\label{fig:rviol}}
\end{figure}

No signal above the expected standard model backgrounds was detected in any of
the topologies investigated, resulting into mass limits or constraints on the 
parameters of the MSSM at least as strong as in the case of R-parity 
conservation, even assuming a dominant $\lambda_{133}$ coupling. This can be
seen, for instance, in Fig.~\ref{fig:rviol} 
for chargino and neutralino searches. 
Interestingly, a substantial region is excluded at LEP~1 beyond the kinematic 
limit for chargino searches at LEP~2. This is due to the large statistics
accumulated at the Z peak and to the fact that the $\Z\to\chi\chi$ decay leads 
to visible final states, in contrast to the case of R-parity conservation. 

The first results of searches for signals of supersymmetry at LEP under the
assumption of a dominant $\lambda'_{ijk}L_iQ_j\overline{D}_k$ coupling were
reported as this chapter was being completed. The final states typically 
involve multijets with leptons or with some missing energy. Here too, the 
kinematic limit for charginos is almost reached~\cite{rviol}.

\section{Outlook}

As indicated in the introduction, it is foreseen that the centre of mass energy 
at LEP~2 will be increased beyond 172~GeV, the energy at which most of the 
results reported above were obtained. Based on the experience already gathered,
it appears that $\sim 10$~\inpb\ are enough to reach the kinematic limit for 
charginos over large regions of the parameter space. Increasing the energy is
therefore always beneficial compared to collecting more luminosity at a given
energy. Once the ultimate LEP~2 energy is reached, it will nevertheless remain
worthwhile accumulating sufficient integrated luminosity to improve the 
sensitivity to sleptons, stops and neutralinos. More optimistically, large
statistics will be welcome to allow the LEP experiments to determine accurately 
the masses and couplings of the various supersymmetric particles which they will
have revealed. 

%\section*{Acknowledgments}
%
%Thanks are due to all my friends for help and support.

\begin{figure}[p]
\begin{center}
\epsfig{file=f1a.eps,width=10cm,
bbllx=0pt,bblly=0pt,bburx=505pt,bbury=505pt}
\end{center}
\begin{center}
\epsfig{file=f1b.eps,width=10cm,
bbllx=0pt,bblly=0pt,bburx=505pt,bbury=505pt}
\end{center}
\caption{(top): A monojet event observed at LEP~1. 
This event is probably due to the
process $\epem\to\Z\gas\to\nunu\qqb$.
(bottom): An acoplanar $\tau$ pair observed at LEP~2. The visible $\tau$ decay
products are a $\rho$ and an a$_1$. This event cannot be interpreted as 
resulting from $\epem\to\WW$ with two $\W\to\tau\nu$ decays. It is rather due 
to $\epem\to\Z\Zs$, with $\Z\to\nunu$ and $\Zs\to\toto$.
\label{fig:monojtoto}}
\end{figure}

\section*{References}

\begin{thebibliography}{99}
\bibitem{PAJ}See for instance: 
P. Janot, {\it ``Searching for Higgs bosons at LEP~1 and LEP~2''},
to appear in Perspectives on Higgs Physics II, World Scientific, ed. G.L. Kane.
\bibitem{EW} The LEP Collaborations, {\it ``A combination of preliminary 
electroweak measurements and constraints on the standard model''}, 
CERN-PPE/96-183.
\bibitem{Warsaw} M. Schmelling and P. Tipton, review talks at ICHEP96, 
Warsaw, July 1996,\hfill\break 
to be published.
\bibitem{monoj} The ALEPH Collaboration, {\it ``Observation of monojet events 
and tentative interpretation''}, 
Phys. Lett. {\bf B 334} (1994) 244.
\bibitem{ochi} The OPAL Collaboration, {\it ``Topological search for the
production of neutralinos and scalar particles''}, 
Phys. Lett. {\bf B 377} (1996) 273.
\bibitem{aslep} The ALEPH Collaboration, {\it ``Search for sleptons in \epem\
collisions at centre-of-mass energies of 161 GeV and 172 GeV''}, 
CERN-PPE/97-056.
\bibitem{AMY} The AMY Collaboration, {\it ``New limits on the masses of the
selectron and photino''},
Phys. Lett. {\bf B 369} (1996) 86.
\bibitem{ostop} The OPAL Collaboration, {\it ``Search for scalar top and scalar
bottom quarks at $\sqrt s =$ 170--172~GeV in \epem\ collisions''},
CERN-PPE/97-046.
\bibitem{ochar} The OPAL Collaboration< {\it ``Search for chargino and 
neutralino production at $\sqrt s =$ 170 and 172~GeV at LEP''},
CERN-PPE/97-083.
\bibitem{achar} The ALEPH Collaboration, {\it ``Searches for charginos and
neutralinos in \epem\ collisions at $\sqrt s =$ 161 and 172~GeV''},\hfill\break 
contribution 614 to EPS-HEP97, Jerusalem, August 1997.
\bibitem{arrgh} The ALEPH Collaboration, {\it ``Update of the mass limit for the
lightest neutralino''}, 
contribution 594 to EPS-HEP97, Jerusalem, August 1997.
\bibitem{avant} The ALEPH Collaboration, {\it ``Mass limit for the lightest
neutralino''},\hfill\break
Z. Phys. {\bf C 72} (1996) 549.
\bibitem{colfac} The ALEPH Collaboration, {\it ``A measurement of the QCD colour
factors and a limit on the light gluino''},
CERN-PPE/97-002.
\bibitem{stable} The ALEPH Collaboration, {\it ``Search for pair-production of
long-lived heavy charged particles in \epem\ annihilation''}, CERN-PPE/97-041;
\hfill\break
The DELPHI Collaboration, {\it ``Search for stable heavy charged particles in
\epem\ collisions at $\sqrt s =$ 130--136, 161 and 172~GeV''},
Phys. Lett. {\bf B 396} (1997) 315.
\bibitem{lphot} The L3 Collaboration, {\it ``Single and multi=photon events with
missing energy in \epem\ collisions at 
161~GeV $<\sqrt s<$ 172~GeV''},
CERN-PPE/97-76.
\bibitem{marie} The ALEPH Collaboration, {\it ``Search for supersymmetric
particles with R-parity violation in Z decays''},
Phys. Lett. {\bf B 349} (1995) 238.
\bibitem{rviol} The ALEPH Collaboration, {\it ``Searches for R-parity violating
supersymmetry at LEP~II''},
contribution 621 to EPS-HEP97, Jerusalem, August 1997.
\end{thebibliography}
\end{document}